\documentclass[12pt]{article}
\usepackage{epsfig,graphics}

\newcommand{\bq}    {\begin{equation}}
\newcommand{\eq}    {\end{equation}}
\newcommand{\bqr} {\begin{eqnarray}}
\newcommand{\eqr} {\end{eqnarray}}
\newcommand{\barr}   {\begin{array}}
\newcommand{\earr}   {\end{array}}
\newcommand{\tc}     {T_{c}}

\newcommand{\tac}    {T < T_{c}}
\newcommand{\tic}    {T > T_{c}}
\newcommand{\sig}    {\sigma}
\newcommand{\siz}    {\sigma_{0}}

\newcommand{\mb}[1]  {\mbox{#1}}

\newcommand{\gev}    {\mb{GeV}}

\newcommand{\dil}    {{\mathrm {dil}}}
\newcommand{\dint}    {{\mathrm d}}
\newcommand{\eint}    {{\mathrm e}}

\begin{document}

\title{Evaporation of the gluon condensate:\\
a model for pure gauge SU(3)$_c$ phase transition}

\author{
\bf Alessandro Drago $^a$, Marina Gibilisco $^b$, Claudia Ratti $^a$\\ 
\\
$^a$ {\small Dipartimento di Fisica, Universit\`a di Ferrara and} \\
 {\small INFN, Sezione di Ferrara, via Paradiso 12, 44100 Ferrara, 
Italy } \\
$^b$ {\small 
ECT*, European Centre for Theoretical Studies in Nuclear Physics} \\
   {\small  and Related Areas, Via delle Tabarelle 286, 
38050 Villazzano (Trento), Italy}
      }


\maketitle

\begin{abstract}

We interpret lattice data for the equation of state of pure gauge 
$SU(3)_c$ by an evaporation model. At low temperatures
gluons are frozen inside the gluon condensate, whose dynamics 
is described in terms of a dilaton lagrangian. Above the
critical temperature quasi-free gluons
evaporate from the condensate: 
a first order transition is obtained by minimizing the thermodynamical
potential of the system.
Within the model it is possible to reproduce lattice QCD results at finite 
temperature for
thermodynamical quantities such as pressure and energy. The gluonic 
longitudinal mass can also be evaluated; it vanishes below the critical 
temperature, where it shows a discontinuity. At very large temperatures we
recover the perturbative scenario and gluons are the only asymptotic degrees of
freedom.

\end{abstract}

\vfill\eject

\section*{}

In recent years, precise data have become available concerning QCD 
thermodynamics at high temperature via numerical simulation on a lattice
(for a recent review, see \cite{Karsch:2001cy}).
Data exist now both for the pure gauge sector and 
for complete QCD at zero
chemical potential; the latter has been explored both in the  
limit of infinite quark masses and in the chiral limit.
Moreover, calculations for finite values of quark masses and for a 
non-vanishing chemical potential are now
appearing; the availability of such a large number of new lattice data
surely represents an important opportunity to test the
effectiveness of models in reproducing the finite temperature
phase transition. 

\smallskip

In this paper, we will concentrate on the simplest case, i.e., the pure
gauge sector. The main known characteristics of 
$SU(3)_{c}$ at finite temperature are the following. 
A first order transition takes place at a temperature
$T_{c}~=~(271\pm 2)~$MeV and at $T\sim$ few $T_{c}$ the asymptotic limit
of a Stefan-Boltzmann gas is not yet reached. A small but not negligible
value for pressure, entropy and energy at $T$ just below $T_{c}$ has been
computed and  the size of the discontinuity of the energy at $T_{c}$,
representing the latent heat of the transition, has also been
estimated.

In our work, we interpret lattice data by assuming a
theoretical scenario similar to the one suggested by Simonov 
\cite{Simonov:1992bc}. In
this approach, at
$T\le T_{c}$ the dynamics of the gluon condensate is dominant, while at $T
> T_{c}$ the condensate evaporates in the form of quasi-free gluons. 
 
Several models have been used to describe 
lattice data (for a review, see \cite{Greiner:1996wv}).
Early attempts were based
on the MIT bag model \cite{Engels:1982ei}, but more sophisticated approaches
became necessary when more precise lattice data started
appearing: for instance, the results for the energy density, the pressure
and the entropy of a pure gluon system 
at $T>T_c$ can be well reproduced by
quasiparticle models in which gluons acquire an effective
temperature-dependent mass 
\cite{Peshier:1994zf,Gorenstein:1995vm,Peshier:1996ty}.
More recently, it has been pointed out that the number of effective
degrees of freedom (i.e. the gluon degeneracy of the system) can itself be
temperature dependent \cite{Levai:1998yx,Schneider:2001nf}.

An interesting idea, similar to the one we will use in this work, 
is implemented in the
so-called ``cut-off'' model 
\cite{Engels:1989ph,Rischke:1992uv,Rischke:1992rk}, 
in which gluons having momenta
smaller than a fixed value $K$ are bound inside non-perturbative
structures and therefore do not directly contribute to the
thermodynamics of the system. 

In the models discussed so far, two important limitations are present:
firstly, the critical temperature $T_{c}$ plays the role of a parameter
and cannot be computed within the model; then, and 
more important, the transition itself is parametrized and not
obtained dynamically, for instance through the minimization of the 
energy of the system.
In the present work we will try to overcome such limitations: in our approach
it is possible to describe the thermodynamical behavior of the
system and obtain a first order transition via the minimization
of the thermodynamical potential; the value of the critical
temperature at which the transition takes place can also be 
estimated~\footnote{A similar approach has been discussed in 
Ref.~\cite{Carter:1998ti},
where the effective degrees of freedom are the dilaton field and constituent
gluons which become massive via an interaction with the gluon condensate. 
However, the results obtained in Ref.~\cite{Carter:1998ti} are not completely 
satisfactory at temperatures just above the critical one.}.   

In our model, we consider three different contributions to the 
thermodynamical potential: the first component comes from the gluon
condensate, whose dynamics is expressed in terms of a dilaton lagrangian
\cite{Schechter:1980ak,Salomone:1981sp,Migdal:1982jp,Ellis:1985jv}; 
then, gluons are introduced in a way similar to the one used in
the cut-off model 
\cite{Engels:1989ph,Rischke:1992uv,Rischke:1992rk}. At variance with
the previous versions of that model, in our case
the cut-off itself is not a
parameter, but rather a function of 
the expectation value of the dilaton field, i.e.
of the gluon condensate. It seems rather natural to assume that, when the
gluon condensate is large, namely at low temperature, many gluons 
are frozen inside this non-perturbative structure: as a consequence, the
infrared cut-off is large. 
On the contrary, when the value of the 
gluon condensate is reduced, gluons having 
large enough momenta may ``evaporate'', and behave as almost-free
particles.  

Finally, the last contribution to the thermodynamical potential is due to
the perturbative gluon-gluon interaction, which is present even at large
temperature, i.e. for 
$T\gg T_{c}$. In the following, we will discuss in sequence 
these three contributions.

\section{Gluon-condensate dynamics}

The idea of a gluon condensate has been introduced many years ago
\cite{Shifman:1979bx,Shifman:1979by}. 
In our approach, we will not try to obtain
a mechanism for gluon condensation, namely a model for the QCD vacuum,
but we will instead describe the dynamics of the gluon condensate
by introducing an effective degree of freedom, i.e. the dilaton field.

In fact, there is a deep connection between the gluon condensation
phenomenon and the violation of the scale invariance, which, in QCD at 
the first loop, is quantified by:
\bq
\langle ~\partial_{\mu}~j^{\mu}_{QCD}~\rangle~=~-{11 N_c\over 96\pi 
^{2}} ~\langle~g^{2}G^{2}~\rangle, \label{1}
\eq
where $j_{QCD}^{\mu}$ is the dilatation current in QCD and $G^2$
is the gluon field strength.
In order to reproduce the QCD scale anomaly and to satisfy low-energy 
theorems \cite{Novikov:1981xj}, a dilaton field has been introduced 
\cite{Migdal:1982jp,Ellis:1985jv}
whose lagrangian reads
\bq
L_{\dil}~=~{1\over 2}(\partial_{\mu}\sigma)^{2}~-~V(\sigma),\label{2}
\eq
where
\bq
V(\sigma)~=~{B\over 4}~\Bigg[\sigma_{0}^{4}~-~\sigma^{4}~+
~4~\sigma^{4}~\ln
\Bigg({\sigma\over \sigma_{0}}\Bigg)\Bigg].\label{3}
\eq
The violation of the scale invariance is given by:
\bq
\partial_{\mu}j_{\dil}^{\mu}~=~4V~-~{\partial V\over
\partial\sigma}\sigma~=~-B\sigma^{4}\label{4}
\eq
and it must satisfy the equality
\bq
\langle ~\partial_{\mu}~j^{\mu}_{\dil}~\rangle~=~
\langle ~\partial_{\mu}~j^{\mu}_{QCD}~\rangle.\label{5}
\eq
The potential $V(\sigma)$ has a minimum at $\sigma=\sigma_{0}$, where 
$V(\sigma_{0})=0$; the small oscillations around the minimum correspond to 
the excitations of a scalar glueball and they can be parametrized as
follows:
\bqr
V(\sigma)&\simeq&2B\sigma_{0}^{2}~(\sigma-\sigma_{0})^{2}~+~
O[(\sigma-\sigma_{0})^{3}] \nonumber \\
&\equiv&{1\over
2}~M_{g}^{2}~(\sigma-\sigma_{0})^{2}~+~O[(\sigma-\sigma_{0})^{3}],\label{6}
\eqr
where the glueball mass $M_{g}$ is
\bq
M_{g}~=~2~\sigma_{0}~\sqrt{B}.\label{7}
\eq
 
The dilaton potential contains two parameters, $\sigma_{0}$ and $B$,
which can be related through eqs. (\ref{1}, \ref{4}, \ref{5}, \ref{7})
to the value of the gluon condensate in the vacuum and to the mass of the
scalar glueball. Concerning the estimate of the gluon condensate 
(for a review, see \cite{Shuryak:1988ck}), 
Ref. \cite{Eidelman:1979xy} indicates the range
\bq
0.12~GeV^{4}~\le\langle(gG)^{2}\rangle~\le0.83~GeV^4,\label{8}
\eq
while both Refs. \cite{Launer:1984ib} and \cite{Novikov:1984jt} 
indicate a value 
$\langle(gG)^2\rangle \sim 0.5$ GeV$^4$ associated, in 
Ref. \cite{Novikov:1984jt}, to an error of about 50\%.

The mass of the scalar glueball has been recently estimated in lattice
QCD, obtaining a mass $M_{g}=(1730\pm 30\pm 80)$ MeV \cite{Morningstar:1999rf}.

The two parameters $\sigma_{0}$ and $B$ are therefore constrained into a
relatively narrow window, the uncertainty being mainly due to the error
bar in the estimate of the gluon condensate.

In the following, we will study the thermodynamical potential associated
with the dilaton lagrangian, at the mean field level, by using the
standard techniques of finite temperature field theory 
\cite{Kapusta:1989,Lebellac:1996}.

The thermodynamics of the dilaton field at finite temperature has 
already been discussed in the literature, for instance in Refs. 
\cite{Agasian:1993fn} and \cite{Carter:1997rf}. In Ref. \cite{Carter:1997rf}
an attempt to go beyond the mean-field
approximation was made, and in Ref.~\cite{Schaefer:2001cn} the gluon condensate
was studied with renormalization group flow equations. We will compare later 
our results with the ones of Ref.~\cite{Schaefer:2001cn}. Due to the 
difficulties associated with the quantization of non-polynomial field theories 
(see e.g.~\cite{Chalmers:1998gn}) we prefer to
stick to the mean-field approximation; moreover, it is not possible to apply 
the renormalization group techniques to our complete model which incorporates 
also gluons evaporating from the condensate.

In the mean field approximation the thermodynamical potential 
reads:
\bq
\Omega_\dil(\sigma,T)~=~V(\sigma)~-~P_{\dil}(\sigma,T),\label{9}
\eq
where $P_{\dil}$, the pressure of the dilaton field, reads\footnote{At very 
large temperatures the contribution of the dilaton field to the thermodynamical
quantities should vanish. To this purpose, in Sec.~\ref{hight} an ultraviolet 
cut-off will be introduced in eq.~(\ref{pdil}).}
\bq
P_{\dil}(\sigma,T)=-T\int{\dint ^{3}p\over
(2\pi)^{3}}~\ln\Big[~1~-~e^{-\omega/T}~\Big],\label{10}
\label{pdil}
\eq
and
\bq
\omega=\sqrt { {\bf p} ^2 + [m(\sigma)]^2 }~ . \label{11}
\eq
The $\sigma-$dependent mass $m(\sigma)$ is defined as 
\bq
m^{2}(\sigma)~=~{\partial^{2} V\over\partial \sigma^{2}}\label{12}
\eq
and it equals $M_{g}^{2}$ for $\sigma=\sigma_{0}$.
The mass squared is negative for $\sigma<{\mathrm e}^{-1/3}\sigma_0$
and therefore the thermodynamical potential gets an imaginary part
for small values of $\sigma$.
There is in the literature a broad discussion about the
physical interpretation of a complex potential and how to deal with it
(see e.g.~\cite{Sollfrank:1995du}). We will follow 
Ref.~\cite{Weinberg:1987vp}, where the imaginary part is interpreted
as a signal of the instability of the system. 
Our recipe for dealing with a complex thermodynamical
potential is therefore all simply to minimize
its $real$ part. Actually the problem
of interpreting the imaginary part of the potential, 
although conceptually important,
turns out to be not so important from a practical viewpoint, since
after the introduction of gluons (what will be done in the next Section) 
the dilaton field will vanish for $\tic$
and we will not have to deal with the instability region.

\begin{figure}

\begin{center}
\includegraphics[width=.7\textwidth]{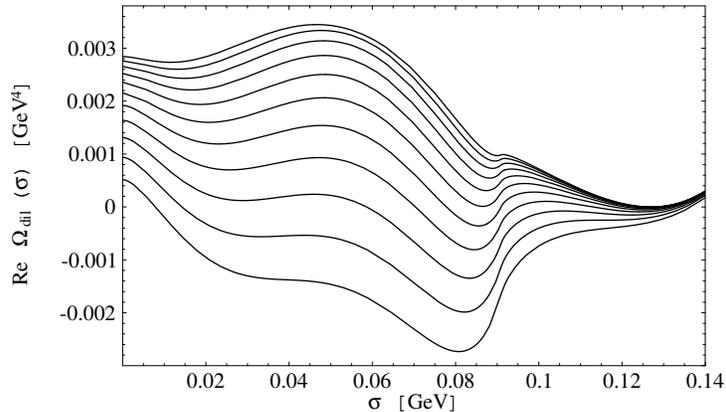}

\caption{
\footnotesize Real part of dilaton thermodynamical potential for various
temperatures. Temperature increases from upper to lower curves,
ranging from 0.2 GeV to 0.4 GeV in steps of 0.02 GeV.
\label{omegadil}}
\end{center}
\end{figure}

In Fig.~\ref{omegadil} we present 
${\mathrm {Re}}\,\Omega_\dil(\sigma,T)$, as a function of $\sigma$,
for various temperatures.
As one can clearly see, when the temperature increases
the real part of the 
thermodynamical potential develops a new minimum for $\sigma
<\sigma_{0}$. At the critical value
$T_{c}$, the new minimum becomes the absolute one and a first order
transition takes place. We must stress again that the transition
of the pure dilaton field appears to be first order due to 
the approximations we have used. The order of the transition
could be established in a consistent way only using a non-perturbative
approach, e.g. studying the dilaton potential on the lattice. 
One should also consider the results of Ref.~\cite{Schaefer:2001cn}, where
no hint of a first order transition was found up to temperatures of the
order of 200 MeV where their prediction reaches its limit of validity.
On the
other hand, many calculations of the behavior of the
dilaton field at finite temperature do indicate a transition
at a temperature similar to the one we get in our approach
\cite{Agasian:1993fn,Carter:1997rf,Sollfrank:1995du}.
It is also interesting to notice that, since in the 
dilaton lagrangian the only dimensional parameter is $\sigma_0$,
the critical temperature as a function of $B$ and $\sigma_0$ must
be of the form:
\bq
T_c^{\dil}=f(B)\, \sigma_0 \, ,
\eq
where $f(B)$ is a function to be determined numerically by minimizing
${\mathrm {Re}}\,\Omega_{\dil}$.

\begin{figure}
\begin{center}
\includegraphics[width= .65\textwidth]{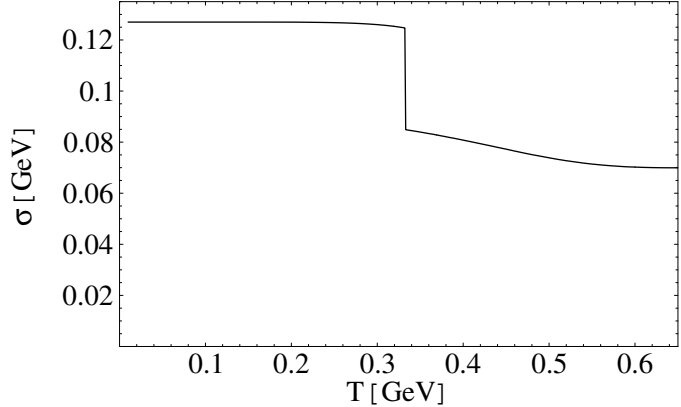}
\end{center}
\parbox{14cm}{
\caption{
\footnotesize Dilaton field as a function of the temperature in the
pure dilatonic model.
\label{dilaton}}
}

\end{figure}

In Fig.\ref{dilaton}, 
we present the expectation value of the dilaton field as a
function of the temperature: for $\tac$ there is a very small reduction of
$\sigma$ from its zero-temperature value $\sigma_{0}$.
This shift of the expectation value of the dilaton field corresponds
to thermal excitations of the glueball. 
At the critical temperature $T_c^{\dil}$, the
dilaton field is discontinuous. Note that for $T>T_c^{\dil}$ the dilaton field
does not vanish. We will see later on that the contribution of 
quasi-free gluons shifts the value of $\sig$ to zero in the deconfined 
phase. Figs.~\ref{omegadil} and \ref{dilaton} 
have been obtained using $B=46.4$ and $\siz=0.127$ GeV,
which correspond to $\langle (gG)^2\rangle=0.35$ GeV$^4$ and
$M_{g}=1.73~\gev$. The critical temperature, in absence of gluons is 
$\tc=\tc^{\dil}=0.3$ GeV: 
the extra pressure of the gluons in the deconfined phase
reduces this value, bringing it close to the one indicated by lattice
calculations.  

\section{The evaporation model}

\subsection{Quasi-free gluons}
In the previous section we have discussed the behavior of the dilaton
field at finite temperature. It is clear that its excitations (i.e.
the glueballs) cannot represent the relevant degrees of freedom at large
temperature, where quasi-free gluons should give the dominant contribution 
to the thermodynamical observables. 
On the other hand, quasi-free gluons should be suppressed below $\tc$,
which, in our model, is the temperature at which the dilaton field is
discontinuous. A simple way to suppress the quasi-free
gluons in the confinement region is by assuming that they are frozen inside the
gluon condensate: when the value of the 
gluon condensate is large, i.e. below $\tc$,
most of the gluons are frozen while, above $\tc$, the condensate
evaporates and gluons become quasi-free particles.
Technically, this idea can be implemented by introducing an infrared cut-off
$K$
in the gluon distribution function, so that only gluons having a momentum
larger than $K$ contribute to the thermodynamics of the system:

\bq
P_{\mathrm {q-free}}(\sigma,T)=-2(N_c^2-1)T\int {\dint ^3 k\over(2\pi)^3}
{\mathrm {ln}}\left [1-\eint^{-k/T}\right ]
\Theta (k-K(\sigma))
\eq
In our model
we assume that the cut-off $K$ is a function of the gluon condensate, 
i.e. of the expectation value of the dilaton field.
The ``evaporation" model has been already discussed in Refs.
\cite{Engels:1989ph,Rischke:1992uv,Rischke:1992rk}, 
but there the cut-off was assumed to be a fixed
parameter. We use for the cut-off the form
\bq
K(\sigma)={A\over \left ({\sigma_0-\sigma \over \sigma_0}\right )^\alpha}\, ,
\eq
so that if $\sigma\rightarrow \sigma_0$, then $K\rightarrow\infty$,
while if $\sigma\rightarrow 0$ then $K\rightarrow A$.
It can be interesting to notice that we can not reproduce
the lattice data satisfactorily if we use a cut-off which
vanishes for $\sigma\ll\sigma_0$. This result seems to indicate
that, even at large temperatures, at which $\sigma\sim 0$, wee gluons
are still suppressed. The value of the parameter $A$ is of the order of
1 GeV and it is therefore natural to interpret this parameter 
as the one regulating
the transition from the perturbative behavior of the gluon
propagator to the non-perturbative one. 

\subsubsection{High temperature degrees of freedom \label{hight}}

At very large temperature the perturbative degrees of freedom should be 
recovered and therefore the dilaton field cannot appear as an effective 
degree of freedom for $T\gg T_c$. On the other hand, at temperatures just
above $T_c$ scalar gluon-gluon correlations described in terms of the 
dilaton field could still be relevant. We will see in Sec.~3 that, above $T_c$,
$\sigma\sim 0$; as a consequence the dilaton
mass vanishes and scalar gluon correlations can 
exist but with a vanishing mass gap~\cite{Agasian:1993fn}.

The idea of describing correlations between the asymptotic degrees of
freedom in terms of effective fields is adopted in many physical problems.
For instance $\sigma$-models have been used in studying the chiral phase 
transition~\cite{Pisarski:1984ms}\footnote{Similarly, in the study of hadronic 
structure, pions 
can be introduced as effective degrees of freedom, whose
substructure can later be resolved in terms of the quark distribution function 
of the pion (see for instance~\cite{Schreiber:1992qx,Suzuki:1998wv}).}. In 
that case, the chiral fields describe 
quark-antiquark correlations in an effective non-perturbative way below and
above $T_c$.
In the scheme we are discussing it is possible, at least in principle, to 
provide a structure
for the dilaton field in terms of gluonic degrees of 
freedom~\cite{Simonov:1992bc}. For simplicity we mimic the dynamics of gluons
inside the dilaton by introducing an ultraviolet cut-off in the dilaton
pressure. Since above $T_c$ gluons having a momentum larger than $K$ are 
quasi-free, and since at least two gluons are needed to produce a scalar 
correlation, we assume that correlations having a momentum larger than $2K$ are
suppressed and do not contribute to the thermodynamical quantities.

We modify therefore eq. (\ref{10}) by introducing an ultraviolet cut-off equal
to $2K$. In this way, at very large temperatures, the dilaton field does not 
contribute, as shown in Fig.~\ref{pparziali5}. On the other hand, the effect of
this cut-off is almost negligible at $T\sim$ a few $T_c$.

\subsection{Residual perturbative interaction}

Lattice
data clearly indicate that even at a temperature larger than $4\,\tc$ 
the Stefan-Boltzmann limit is not yet reached and the data for pressure,
energy and entropy lie below the free-gas limit. As we will see, the
introduction of an infrared cut-off is not sufficient to explain both the
data near $\tc$ and those at large temperatures. 
It is therefore necessary to
introduce perturbative corrections, which have a relevant role for
$\tic$. These corrections are well established and a huge amount of
work has been devoted to their estimate (for a recent review
see \cite{Blaizot:2001nr}). In the present work we are not interested in
a detailed comparison with the data at $T\gg T_c$, but we mainly aim 
at describing the data near the critical temperature. We have therefore
adopted the simplest prescription, which 
consists in taking into account only the first order $O(g^2)$ corrections.
These corrections have been computed in Ref.\cite{Kapusta:1979fh},
but in the present case additional $\Theta$-functions occur which
restrict the phase space for the interacting gluons 
(see \cite{Greiner:1996wv}). The
$O(g^2)$ contribution to the pressure reads:

\bqr
&&P_{\mathrm {int}}(\sigma,T)=g^2 N_c(N_c^2-1)\Bigg\{
-3 \left(\int {\dint ^3 k\over(2\pi)^3}
{1\over k}N_B\left({k\over T}\right)
\Theta [k-K(\sigma)]\right)^2\nonumber \\
&&\,\,+ \int {\dint ^3 k_1\over(2\pi)^3} \int {\dint ^3 k_2\over(2\pi)^3}
\,{1\over k_1 k_2}N_B\left({k_1\over T}\right) N_B\left({k_2\over T}\right)
\Theta [k_1-K(\sigma)]\, \Theta [k_2-K(\sigma)]
\nonumber\\
&&\qquad\times\left ({9\over 4}\, 
\Theta [\vert {\mathbf k_1}+{\mathbf k_2}\vert - K(\sigma)]
-{1\over 4}\,
\Theta [\vert {\mathbf k_1}-{\mathbf k_2}\vert - K(\sigma)]\right )
\Bigg\} \,\, ,
\eqr
where $N_B(x)=({\mathrm e}^x-1)^{-1}$ is the Bose-Einstein distribution.
Here $g^2$ is the temperature-dependent running coupling-constant
\bq
g^2(T)={48\pi^2\over 11 N_c {\mathrm {ln}}\,[(T^2+S^2)/\Lambda^2]} \,\, ,
\eq
where we have introduced a regulator $S$, whose 
appearance can be related to the existence of a minimal
momentum $K$ for the propagating gluons \cite{Greiner:1996wv}\footnote{In 
principle, $g^2$ can also depend on the value of the gluon 
condensate~\cite{Adler:1984zh}. This possibility will be explored in a future
work~\cite{tutti}.\label{nota}}. 
As we shall see, the perturbative
corrections play a relatively minor role in our model and more
sophisticated choices of the running coupling would hardly affect the
results. The typical value for $S$ is $S\sim$ GeV.  

\vfill
\newpage

\section{Results}

The parameters in our model are the following:
\begin{itemize}
\item
dilaton lagrangian: $B$, $\sigma_0$
\item
quasi-free gluons: $A$, $\alpha$
\item
running coupling-constant: $\Lambda$, $S$.
\end{itemize}
Concerning $B$ and $\sigma_0$, as discussed in Sec.~1 their value is
bounded by the ``experimental'' value of the gluon condensate and by
the lattice result for the mass of the scalar glueball $M_g$. In the
following we will present results obtained using $B=46.4$ and 
$\sigma_0=0.127$ GeV, which correspond to $\langle(gG)^{2}\rangle=0.35$ GeV$^4$
and to $M_g=1.73$ GeV, both near the center values indicated for
these quantities.
Concerning the quasi-free gluons, we use $A=1.015$ GeV and $\alpha=0.5$.
Finally, the parameters for the running coupling-constant are not too strictly
constrained in our calculation, since the perturbative interaction
turns out to play a minor role in our model. 
We adopted $S=7.15$ GeV and $\Lambda=T_c$, where $T_c$ is the value of the
critical temperature computed in the model. These parameter values are also 
consistent with the more general form for $g^2$ introduced in~\cite{tutti} 
(see note~\ref{nota}). 
Notice that the value of the
critical temperature is not modified by the perturbative corrections
and can therefore be computed before the latter are taken into account.
\subsection{Thermodynamical quantities}

The results we present are obtained by minimizing the real part of 
the total thermodynamical potential 
$\Omega_{\mathrm {tot}}(\sigma,T)$, as a function of $\sigma$,
for a given temperature $T$, where:
\bq
\Omega_{\mathrm {tot}}(\sigma,T)=\Omega_\dil(\sigma,T)-
P_{\mathrm {q-free}}(\sigma,T)-
P_{\mathrm {int}}(\sigma,T)\,\, .
\eq 

\begin{figure}
\begin{center}
\includegraphics*[width= .7\textwidth]{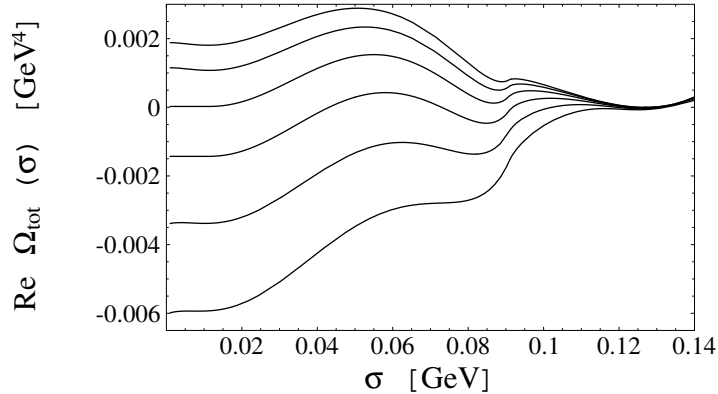}
\end{center}
\parbox{14cm}{
\caption{
\footnotesize Real part of total thermodynamical potential for various
temperatures. Temperature increases from upper to lower curves,
ranging from 0.23 GeV to 0.33 GeV in steps of 0.02 GeV.
\label{omegatot}}
}

\end{figure}

In Fig.~\ref{omegatot} we present the real part of the thermodynamical
potential. Due to the extra
pressure of quasi-free gluons, 
a new minimum develops for $\sigma = 0$. In this way we
avoid entering the region of instability of the dilaton field, where
imaginary parts develop, since the expectation value of the field 
jumps from a value  
$\sigma\sim\sigma_0$ to $\sigma=0$.  
The value of the critical temperature, which corresponds to the discontinuity
of the dilaton field, is $T_c=0.27$ GeV.  The reduction of the value of the
critical temperature, due to the introduction of gluons in the deconfined
phase, is rather independent of the precise value of the 
cut-off parameters and it turns out to be always of the order of
10\%. The critical temperature $T_c^\dil$ is therefore a rather
good approximation to the value of $T_c$ as computed from the
complete model.

\begin{figure}
\begin{center}
\includegraphics[width= .7\textwidth]{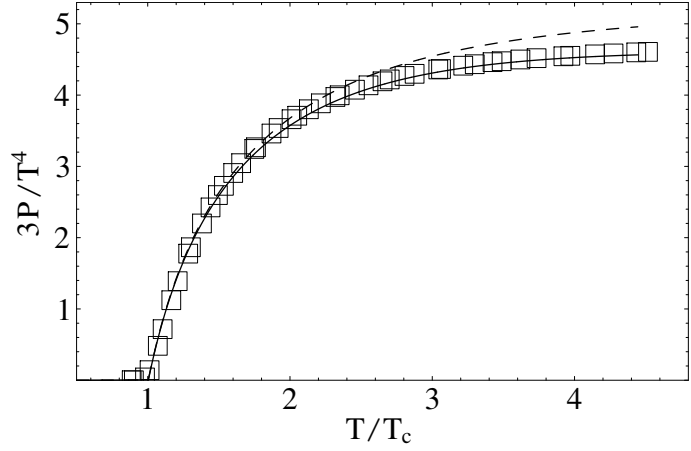}
\end{center}
\parbox{14cm}{
\caption{
\footnotesize Scaled pressure in our model compared with
lattice data. The solid line is obtained minimizing the
total thermodynamical potential, while the dashed line is obtained
neglecting $P_{\mathrm {int}}$.
\label{presstot}}
}

\end{figure}

In Fig.~\ref{presstot} we compare our result for the
scaled pressure with the lattice data \cite{Karsch:2001cy}.
The pressure is obviously connected to
the thermodynamical potential by the relation
$P=-\Omega_{\mathrm {tot}}$.
We also show the pressure obtained minimizing the 
thermodynamical potential neglecting the interaction term $P_{\mathrm {int}}$.
It is clear that the interaction modifies significantly the pressure only
for temperatures of the order of $2\, T_c$, or larger. The critical 
temperature,
as well as the shape of the pressure near $T_c$,
are independent of the interaction contribution. 
As already stated, we are not particularly interested 
in reproducing in a very accurate way the data at large
temperatures, introducing perturbative corrections in a sophisticated way,
our main aim being to describe the phase transition.

\begin{figure}

\begin{center}
\includegraphics[width= .7\textwidth]{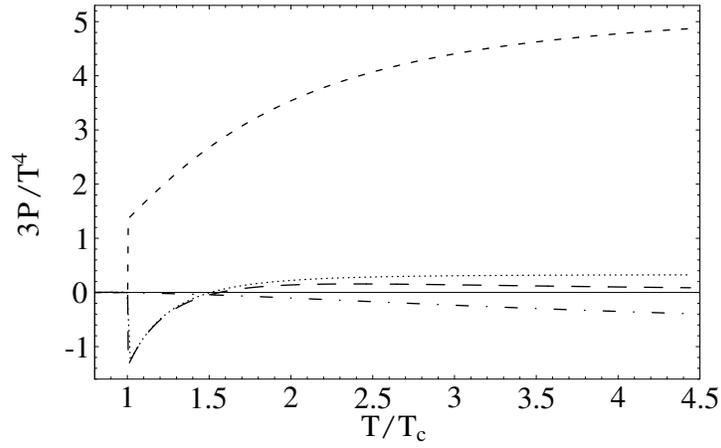}
\end{center}

\parbox{14cm}{
\caption{
\footnotesize Various contributions to the total pressure. The long-dashed line
corresponds to $P_\dil=-\Omega_\dil$, the 
short-dashed line is $P_{\mathrm {q-free}}$, the dashed-dotted line is
$P_{\mathrm {int}}$ and the dotted line is $P_{\mathrm {dil}}$ but without 
the ultraviolet cut-off.
\label{pparziali5}}
}

\end{figure}

In Fig.~\ref{pparziali5} we present the decomposition of the total pressure
into its various contributions. As it can be seen, the dilaton
contribution and the quasi-free-gluons one are both discontinuous
at the critical temperature, but their discontinuities cancel
so that the total pressure is continuous. In Ref.\cite{Simonov:1992bc},
such a behavior for the gluon condensate and for the quasi-free gluons
was anticipated. From the figure it is also clear that the
perturbative correction $P_{\mathrm {int}}$ vanishes at $T_c$.
Let us remark again that the dilaton gives a 
contribution to the scaled pressure which vanishes at large temperatures, due 
to the presence of the ultraviolet cut-off. 

\begin{figure}
\begin{center}
\includegraphics[width= .7\textwidth]{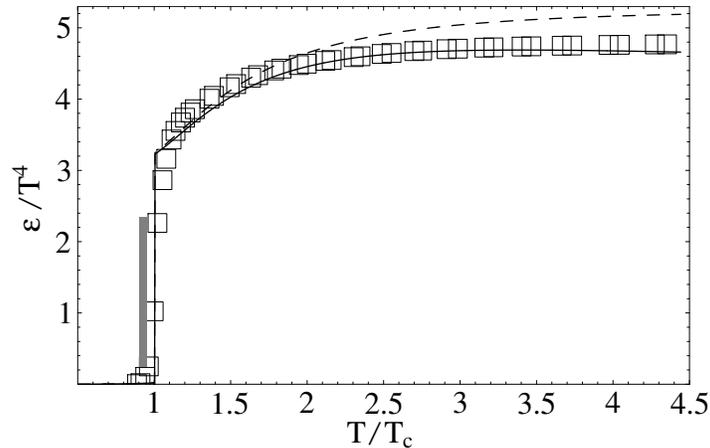}
\end{center}
\parbox{14cm}{
\caption{
\footnotesize Scaled energy density in our model compared with
lattice data. The solid line is obtained minimizing the
total thermodynamical potential, while the dashed one is obtained
neglecting $P_{\mathrm {int}}$. The shaded rectangle indicates
the latent heat as estimated in lattice calculations.
\label{energy}}
}

\end{figure}

In Fig.~\ref{energy} we compare the energy density
$\epsilon=T {\dint P\over\dint T}-P$ computed in our model with
lattice data\footnote{We computed also the entropy density in our model.
Considerations similar to the ones done for the energy density
can be done for the entropy.}. 
We also show the lattice result for the latent heat.
The main difference between our result and the lattice one is that
at $T<T_c$ our energy density is considerably smaller than the one
indicated by lattice calculations. This discrepancy is due to the
presence, in our calculation, of the scalar glueball only, while
the $J=2$ glueball should also contribute. The mass of the latter
has been estimated to be $M_{2^{++}}=2400\pm 25\pm 120$ 
\cite{ Morningstar:1999rf}. The introduction of these new degrees of freedom
would correspond, roughly, to a degeneracy factor of 6 in front of
the glueball contribution \cite{Kapusta:1989}, and would bring
the computed energy density close to the lattice one
at $T<T_c$. For simplicity we have not included the excitations
of the tensor glueball in our calculation, but it can obviously be done
in the future.

\begin{figure}
\begin{center}
\includegraphics[width= .85\textwidth]{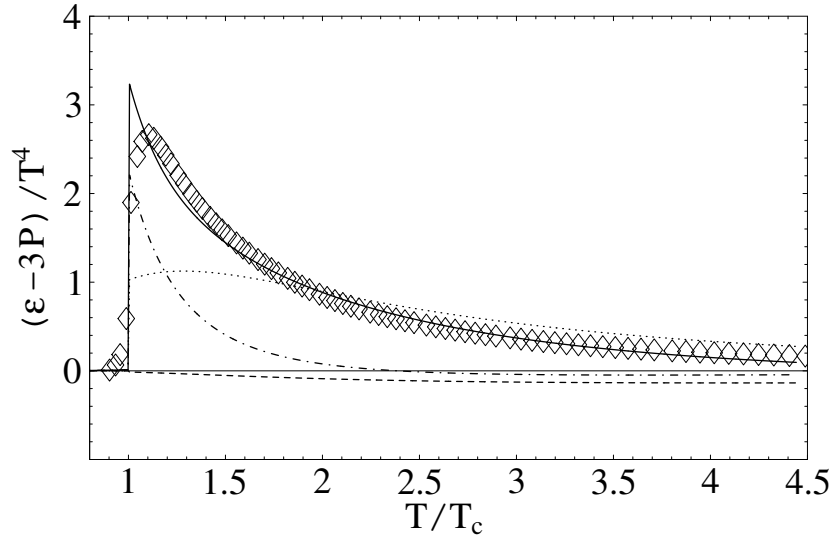}
\end{center}
\parbox{14cm}{
\caption{
\footnotesize Interaction measure. The solid line is the total value,
the dot-dashed is the dilaton contribution, the dotted is
the one due to quasi-free gluons and the dashed corresponds
to the perturbative interaction.
\label{interaction}}
}

\end{figure}

We must also notice that lattice data for the energy decrease much
faster than the results of our model for temperatures just above $T_c$. 
We have found that, using other parameter sets, it is possible in principle
to reduce this discrepancy. However, in those cases the dilaton field does
not jump directly to zero at $T_c$, but it reaches a small but finite value.
Due to the difficulties associated with the mean field treatment of the model
for small values of $\sigma$ we have not explored this possibility in detail. 

Finally, in Fig.~\ref{interaction} we show the various contributions
to the ``interaction measure'', namely the (scaled) quantity
indicating the distance from the Stefan-Boltzmann relation 
$\epsilon=3 P$. For $T\sim T_c$ the main contribution to this
``interaction measure'' comes from the dilaton field. The
contribution due the quasi-free gluons
is large at moderate temperature, but it decreases more rapidly than
the contribution due to $P_{\mathrm {int}}$, which is the dominant
term at very high temperature. 

\vfill
\newpage

\subsection{Thermal gluon masses}
In a covariant gauge, the gluon propagator can be written in the following 
general form
\bq
D_{\mu\nu}=\frac{1}{F-q^2}P_{L}^{\mu\nu}+\frac{1}{G-q^2}
P_{T}^{\mu\nu}+\frac{\xi q_{\mu}q_{\nu}}{\left({\bf q}^2\right)^2}
\eq
where $P_{L}^{\mu\nu}$ and $P_{T}^{\mu\nu}$ are the longitudinal and transverse
projection operators, defined as
\bqr
P_{T}^{00}&=&P_{T}^{0i}=P_{T}^{i0}=0
\nonumber\\
P_{T}^{ij}&=&\delta^{ij}-q^iq^j/{\bf q}^2
\nonumber\\
P_{L}^{\mu\nu}&=&q^{\mu}q^{\nu}/q^2-g^{\mu\nu}-P_{T}^{\mu\nu}.
\label{proiettori}
\eqr
The gluon self-energy is given by
\bq
\Pi^{\mu\nu}=G P_{T}^{\mu\nu}+F P_{L}^{\mu\nu}, 
\label{pai}
\eq
$G$ and $F$ being scalar functions of $q^0$ and $|\bf{q}|$. The electric and 
magnetic masses are defined as
\bqr
F(0,{\bf q}\rightarrow 0)&=&-\Pi_{00}(0,{\bf q}\rightarrow 0)=m^{2}_{el}
\nonumber\\
G(0,{\bf q}\rightarrow 0)&=&\frac 12\Pi_{ii}(0,{\bf q}\rightarrow 0)=
m^{2}_{mag},
\eqr
where the relation between $F$, $G$ and $\Pi$ comes from 
eqs.~(\ref{proiettori}) and (\ref{pai}).

\begin{figure}
\parbox{6cm}{
\scalebox{1.3}{
\includegraphics*[55,564][377,738]{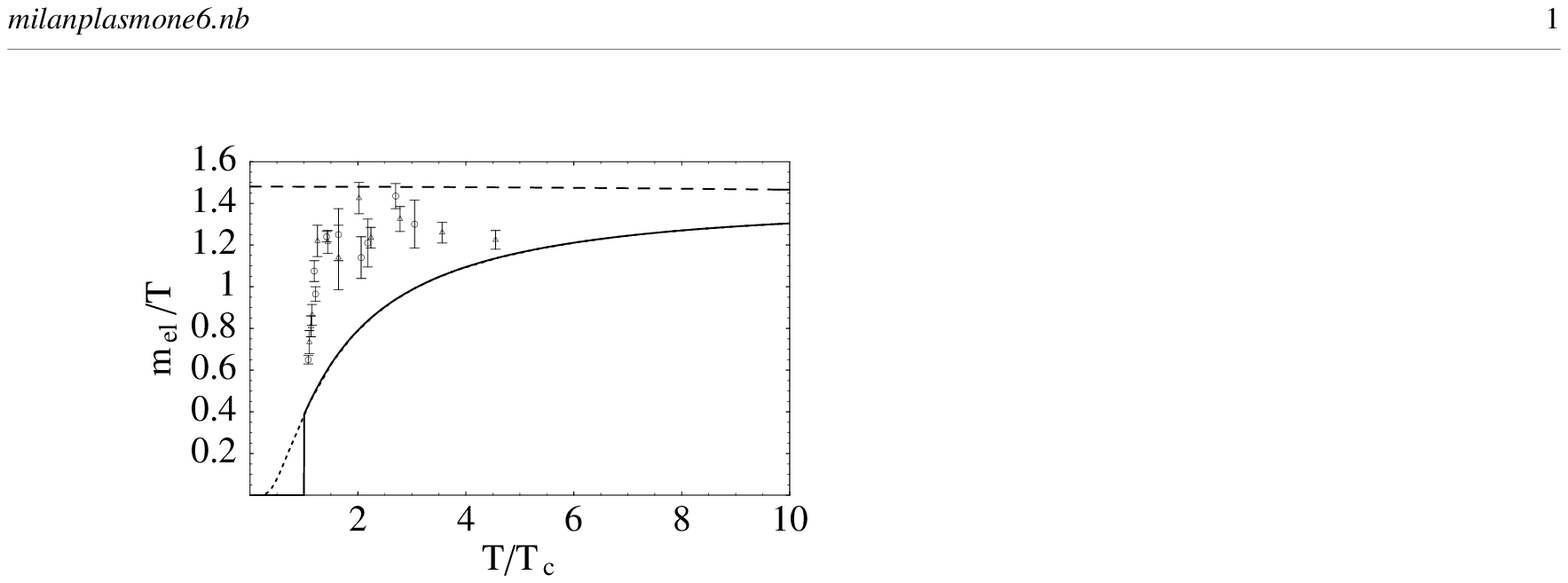}}
}

\parbox{14cm}{
\caption{
\footnotesize Gluonic electric mass in our model (solid line), and in a
purely perturbative calculation without cut-off (dashed line). The data, taken 
from Ref.~\cite{Kaczmarek:1999mm}, correspond to half the screening mass 
$\mu(T)/2$ (see text). The dotted line indicates $m_{el}$ in the metastable 
phase.
\label{mgluon}}
}

\end{figure}

The gluon self-energy can be evaluated perturbatively;
we will concentrate on the electric mass, due to the 
difficulties associated with the magnetic one.
Two different contributions arise for $\Pi^{00}$, a zero temperature one, which
vanishes after renormalization, and a finite temperature one, which reads
\bq
\Pi^{00}=-12 g^2 \int\frac{{\mathrm d}^3 p}{(2\pi)^3}\frac{1}{\omega}\frac{1}
{\eint^{\beta\omega}-1}~~.
\eq
In our model, which allows contributions to the thermodynamics of the system
only from gluons having a momentum larger than $K$, we rewrite the
above formula as:
\bq
\Pi^{00}=-12 g^2 \int\frac{{\mathrm d}^3 p}{(2\pi)^3}\frac{1}{\omega}\frac{1}
{\eint^{\beta\omega}-1}\theta\left(p-K\left(\sigma\right)\right).
\eq
In Fig.~\ref{mgluon} we show our results;
the gluonic electric mass presents a discontinuity
at $T=T_c$, because of the discontinuity of the $\sigma$ field. 
We also show (dotted line) $m_{el}$ in the supercooled, 
metastable phase in which $\sigma$ 
is kept
equal to $\sigma(T=T_{c}^{+})$ for $T<T_c$ (see page. 152 of 
Ref.~\cite{Kapusta:1989}).
For comparison we present the electric mass in a standard perturbative 
calculation
without the cut-off, which yields $m_{el}=gT$ (dashed line).

To compare our results with lattice QCD calculations we must recall that a 
screening mass $\mu(T)$ can also be introduced by considering the behavior of 
the potential $V(R,T)$ between gauge invariant sources at $T>T_c$. 
This potential can be parametrized as~\cite{Kaczmarek:1999mm}:
\bq
\frac{V(R,T)}{T}=\frac{e(T)}{(RT)^d}\eint^{-\mu(T)R}
\eq
where $d$, $e(T)$ and $\mu(T)$ are determined from lattice results.
In perturbative calculations, the screening mass $\mu(T)$ and the gluon
electric mass $m_{el}$ are connected by the simple relation
\bq
\mu=2m_{el}\,\, .
\label{masses}
\eq
In Fig.~\ref{mgluon} we see that this perturbative relation is recovered for
$T\gg T_c$. For temperatures just above $T_c$ the situation
is less clear, also due to the large error bars in lattice data,
but the simple proportionality indicated by eq.~(\ref{masses}) seems
to be (slightly) violated.

Concerning the magnetic mass, it is well known that it vanishes
in a perturbative calculation. Notice anyway that, due to the
infrared cutoff which characterizes the evaporation model, 
higher order loops are not expected to be affected by 
infrared divergences,
even in the absence of a magnetic mass.
For the same reason, in a cut-off model like the one we have discussed here,
perturbative corrections can be computed in principle up to an arbitrary order;
moreover, all these contributions vanish at $T=T_{c}^{+}$.

\section{Conclusions}
We have presented a simple realization of an evaporation model,
in which at low temperature gluons are frozen inside the 
non-perturbative condensate while at high temperature they
escape from the condensate and behave as quasi-free particles. 
We have shown that it is possible within the model to reproduce the main
results of lattice QCD for thermodynamical quantities such as pressure and
energy. At variance with other models for these 
quantities, in our approach a first order transition is obtained
by minimizing the thermodynamical potential and the latent heat can
be estimated. It is also possible
within the model to study finite temperature gluon masses. Here
again our results are consistent with the indications of lattice QCD.

The main problem in the present analysis arises from the
difficulties associated with the appearance of an imaginary part 
in the thermodynamical potential. We assumed that the imaginary part
signals an instability of the system and we have therefore minimized
the real part of the thermodynamical potential. This difficulty is clearly
related to the mean-field approximation we have adopted. Although
this technique seems sufficient to reproduce rather precisely
lattice QCD results, it is clear that only more sophisticated
approximations can clarify the details of the behavior of the
system near the critical temperature. However, in the complete model,
which includes gluons in the deconfined phase, the expectation
value of the dilaton field is such that the system does not enter
the unstable region at any temperature.

This calculation can be extended in several directions.
Firstly, it will be important and, probably,
relatively easy to study the behavior of the
phase transition as a function of the color number $N_c$. Lattice calculations
discussing the dependence on $N_c$ of the critical temperature, 
of the glueball masses and of the gluon condensate appeared recently
(see for instance Refs.~\cite{Lucini:2001ej,Lucini:2001nv}),
allowing comparisons with our model. 
As pointed out in~\cite{Simonov:1992bc}, in this model
the value of the critical
temperature is $N_c$ independent in the large
$N_c$ limit.

Another, more important extension
of the present work will be the inclusion of quarks. This
can be done by dressing the quark propagator via the Schwinger-Dyson
equation and by using a quark-gluon coupling which depends on the value
of the gluon condensate. Work along these
lines is now in progress \cite{tutti}.

\bigskip
\bigskip
\bigskip
\noindent
It is a pleasure to thank V.~Barone, T.~Calarco, M.~Caselle, P.~Castorina,
M.P.~Lombardo, R.~Tripiccione for
many useful discussions and F.~Karsch for providing us the most
recent lattice data. We would like also to thank ECT* for
support and hospitality.

\bigskip
\bigskip

\bibliography{biblio3}
\bibliographystyle{h-physrev3}

\end{document}